\documentclass[letterpaper, 10 pt, conference]{ieeeconf}  

\IEEEoverridecommandlockouts                              

\usepackage{graphicx}
\usepackage{amsmath}
\usepackage{amsfonts}
\usepackage{amssymb}
\usepackage{float}
\usepackage{color}
\usepackage{subcaption}
\usepackage{caption}
\usepackage{balance}
\usepackage{cite}
\usepackage{multirow}
\usepackage{array}
\usepackage{booktabs,array,dcolumn}
\newcolumntype{P}[1]{>{\centering\arraybackslash}p{#1}}
\usepackage{algorithm}
\usepackage{cite}
\usepackage{color}
\usepackage{bm}
\usepackage{amsmath}
\usepackage{graphicx}
\usepackage{float}
\usepackage{enumerate}
\usepackage{color}
\usepackage{cite}
\usepackage{amsmath} 
\usepackage{amssymb}  
\usepackage{bm}
\usepackage{balance}
\title{\LARGE \bf
A Cyber-Security Investment Game for Networked Control Systems
}

\author{Pratishtha Shukla$^{1,2}$, Aranya Chakrabortty$^{1,3}$, and Alexandra Duel-Hallen$^{1,4}$\\
$^{1}$Electrical \& Computer Engineering, North Carolina State University\\
Email: $^{2}$pshukla@ncsu.edu, $^{3}$achakra2@ncsu.edu, $^{4}$sasha@ncsu.edu\\
\thanks{This research is partly supported by the US National Science Foundation under grant ECCS 1544871.}}

\begin{document}

\maketitle
\thispagestyle{empty}
\pagestyle{empty}

\begin{abstract}

We formulate a resource-planning game between an attacker and a defender of a network control system. We consider the network to be operating in closed-loop with a linear quadratic regulator (LQR). We construct a general-sum, two-player, mixed strategy game, where the attacker attempts to destroy communication equipment of some nodes, and thereby render the LQR feedback gain matrix to be sparse, leading to degradation of closed-loop performance. The defender, on the other hand, aims to prevent this loss. Both players trade their control performance objectives for the cost of their actions. A Mixed Strategy Nash Equilibrium (MSNE) of the game represents the allocation of the players' respective resources for attacking or protecting the network nodes. We analyze the dependence of a MSNE on the relative budgets of the players as well as on the important network nodes that must be preserved to achieve a desirable LQR performance. MSNE is computed using nonlinear programming. Results are validated using the New England power system model, and it is shown that reliable defense is feasible unless the cost of attack is very low or much smaller than the cost of protection per generator. 

\end{abstract}
\vspace{0.15cm}

\section{Introduction}
Cyber-physical security of network control systems is a critical challenge for the modern society. 
Recently there has been a renewed surge of interest on this topic in the control systems community where researchers are trying to understand how cyber-physical attacks can jeopardize closed-loop stability and performance of network control systems.
Several illustrative results, tutorials, and surveys on this topic have been presented in recent papers such as \cite{mo,tabuada,sastry,pasqual3}. 
Game theory is a common tool for modeling and analyzing cyber-security problems as it effectively captures the conflict of goals between attackers and defenders. The attacker tries to maximize the damage to the closed-loop system while the defender tries to minimize it. 
For example, \cite{YuanYuan} studies control under denial-of-service (DoS) attacks using a unified game approach, and derive optimality criteria and algorithms for attackers versus defenders using optimal control theory. Similarly, in \cite{Gupta} a dynamic multi-player non-zero sum game with asymmetric information and resource constraints is developed to obtain equilibrium solution under DoS attacks. The majority of these investigations, however, are focused on data manipulation or DoS attacks; the topic of malicious destruction of communication hardware (e.g. circuit boards, memory units, and communication ports), in contrast, has received relatively little attention. In reality, hardware attacks can cause much more serious and lasting degradation 
\cite{mclaughlin} since a single hardware failure can disable communications or cause persistent jamming for an extended period of time and require extensive repairs, as opposed to intermittent data interruptions associated with DoS or detectable data integrity attacks \cite{detection-pasqual}.

A legitimate question, therefore, is how can network operators invest money for securing the important assets of their system against hardware attacks under a limited budget? The same question applies to attackers in terms of targeting the best set of hardware devices to maximize damage under a limited budget. These types of investment games for cyber-security 
are still open research questions. Some preliminary results have been reported in \cite{Sarabi,Hota,Amin}. In \cite{Sarabi}, a strategic one-shot investment game with perfect information between an attacker and defender is proposed for a multi-node network, where the loss in the network graph is proportional to the number of nodes under successful attack. A game-theoretic framework to compute optimal and strategic security investments by multiple defenders is introduced in \cite{Hota} using inter-dependency graphs. Neither of these two papers, however, consider any control-theoretic objective. In \cite{Amin} the gap between the individual and socially optimal security investment is investigated for a network of identical control systems. While a Linear Quadratic Gaussian (LQG) control objective is used within each sub-system, the security interdependency model among the sub-systems is not directly related to the physical system model.

In this paper, we develop a strategic game between an attacker and a defender in a network control system with linear quadratic regulator (LQR) based state-feedback control $\bm{u}(t)=-\textbf{K}\bm{x}(t)$. The need for feedback control in our model guides the level of attack and security investment.  We consider hardware-level attacks that disable all communication associated with a subset of network nodes, including links for self-feedback, and thereby induce sparsity in the feedback gain matrix $\textbf{K}$. The actions of the players are given by a set of possible sparsity patterns \cite{ss_Mihailo} in $\textbf{K}$. The attacker invests her resources to promote these patterns in such a way that the value of the closed-loop LQR objective function increases sharply while the defender invests in tamper-resistant devices \cite{mo}, intrusion monitoring, devices or software that ensure authorized and authenticated access via increased surveillance \cite{enisa},
etc., in order to prevent such harmful sparsity patterns in $\textbf{K}$. A resulting Mixed Strategy Nash Equilibrium (MSNE) describes the resource allocation of the two players over a range of cost parameters, reflecting their relative budgets. As the cost of defense per node increases, MSNE reveals the identity of the important physical nodes \cite{Mihailo} that need to be protected. We validate our results using the IEEE New England power system model and determine the defender's budget requirements to ensure successful protection of closed-loop LQR performance. 

The rest of the paper is organized as follows. In Section II, we present the system model. In Section III, we formulate the proposed mixed-strategy investment game for network control systems. Numerical results and performance evaluation for the New England power grid model are provided in Section IV. Section V concludes the paper.

\section{System Model}
We consider a Network Control System (NCS) with $n$ nodes, where each node may contain multiple states and control inputs, as shown in Fig. 1. The state vector $\bm{x}_i \in \mathbb R^{m_i}$ measured at the $i^{th}$ node, $i=1,\dots,\,n$ is communicated to the node itself as well as to other nodes for taking control actions. We refer to the self-communication links as {\it local} or {\it intra-node} links, and to those between the nodes as {\it inter-node} links. The state-space model of the network is written as 
\vspace{-0.08cm}
\begin{equation} \label{1}\tag{1}
\bm{\dot{x}}(t) = \bm{Ax}(t) + \bm{Bu}(t) +\bm{D}w(t),\quad \bm{x}(0) = \bm{x}_{0}.
\end{equation}
where $\bm{x}(t)=(\bm{x}_1^T(t),...,\bm{x}_n^T(t))^T \in \mathbb{R}^{m\times 1}$ is the vector of states, $\bm{x}_j \in \mathbb{R}^{m_j \times 1}$ is the vector of states for node $j\in \left\{1,...,n\right\}$, $\bm{u}(t)=(\bm{u}_1^T(t),...,\bm{u}_n^T(t))^T \in \mathbb{R}^{r\times 1}$ is the vector of control inputs, $\bm{u}_j \in \mathbb{R}^{r_j \times 1}$ is the vector of control inputs of node $j$, $w(t)$ is a scalar impulse disturbance, $\textbf{A} \in \mathbb{R}^{m \times m}$, $\textbf{B} \in \mathbb{R}^{m \times r}$, $\textbf{D} \in \mathbb {R}^{m \times 1}$ are the state matrix, input matrix, and disturbance matrix, respectively. The physical topology of the network is assumed to be reflected in $\textbf{A}$. The controller $\bm{u}(t)$ is assumed to be a linear state-feedback control of the form 
\vspace{-0.2cm}
\[\bm{u}(t)=-\textbf{K}\bm{x}(t),\tag{2}\label{2}\]
where $\textbf{K} \in \mathbb{R}^{r\times m}$ is the feedback gain matrix is given by
\begin{equation}
\tag{3}\label{3}
\textbf{K}= \begin{bmatrix}
\textbf{K}_{11} & \textbf{K}_{12} & \cdots & \textbf{K}_{1n}\\
\textbf{K}_{21} & \textbf{K}_{22} & \cdots & \textbf{K}_{2n}\\
\vdots & \vdots & \cdots & \vdots\\
\textbf{K}_{n1} & \textbf{K}_{n2} & \cdots & \textbf{K}_{nn}\\
\end{bmatrix}
.
\end{equation}
Here, the block $\textbf{K}_{ij}$ represents the feedback gains from the states of node $j$ and the controller of node $i$. The control objective is to minimize the quadratic function 
\[J(\textbf{K})=\int_{t=0}^{\infty} [\bm{x}(t)^{T}\textbf{Q}\bm{x}(t)+\bm{u}(t)^{T}\textbf{R}\bm{u}(t)] dt,\tag{4}\label{4}\]
where $\textbf{Q}=\textbf{Q}^T \succeq 0 \in \mathbb{R}^{m \times m}$ and $\textbf{R}=\textbf{R}^T \succ 0 \in \mathbb{R}^{r \times r}$ are given positive semi-definite and positive definite design matrices, respectively. 
\begin{figure}[t]
\centering
\includegraphics[scale=0.55]{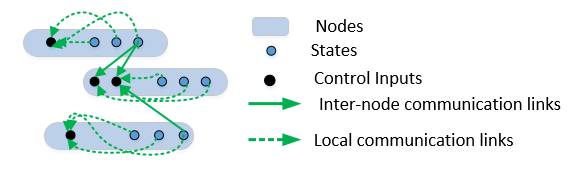}
\caption{\small{System topology}}
\end{figure}

\section{Mixed Strategy Investment Game}
We consider hardware-level attacks that aim to disable or persistently jam the communication within and to/from a subset of nodes of a NCS, thus preventing feedback of states from and reception of control inputs into these nodes, degrading the system-wide control performance. To model the interaction between the attacker and a defender, we employ a strategic, two-player, mixed-strategy (MS) \cite{Wiley-NLP} investment game. In the proposed game, the attacker invests into disabling the communication equipment while the defender invests into protecting it. The players have opposite control performance goals \textemdash the attacker aims to increase the expected system-wide energy (4) while the defender tries to keep it as low as possible. The player's action sets and the control performance objectives are outlined in Section III.A. Moreover, the players aim to reduce their costs of attack/protection, resulting in combined objectives (payoffs), which include both the control performance metric and the cost for each player's action as detailed in Section III.B. In the proposed game, the players choose from the subsets of nodes to attack/protect according to their mixed strategies, i.e., the probabilities assigned to each subset. The game solution is given by a MSNE as detailed in Section III.C. The outcome of the game is a stochastic investment strategy, which jointly optimizes the cost/performance trade-offs for both players. The overall objective is to determine the regions of players' relative budgets where cyber-security protection is effective.

\subsection{Player's Actions and the Energy Loss}
The action spaces of both players are given by the set $\bm{\mathcal{X}}$ of all possible binary \textit{N}-tuples
\begin{equation*}
\label{5}\tag{5}
\begin{split}
\bm{\mathcal{X}} = \left\{\textbf{x}_{\textit{m}} = (x_{\textit{m}}^{\textit{1}},...,x_{\textit{m}}^{\textit{n}}) \ | \ x_{\textit{m}}^{\textit{p}} \in \left\{0,1\right\}, \right. \\ 
p =1,...,n, \ m = 0,...,N-1, \\
\left. N = 2^n \right\}.
\end{split}
\end{equation*}

We refer to the $N$ elements $\textbf{x}_m$ of $\bm{\mathcal{X}}$ as {\it sparsity patterns}. Possible actions of the attacker are the ``attack" sparsity patterns $\bm{a}_i \in \bm{\mathcal{X}}$, where $a_i^k = 0$ implies the $k^{th}$ node is attacked (i.e., the attacker attempts to disable all communication/feedback within and to/from node $k$), and $a_i^k = 1$ means the communication of node $k$ remains intact. Similarly, the actions of the defender are all ``protection" patterns $\bm{p}_j \in \bm{\mathcal{X}}$, where $p_j^k = 1$ if the defender protects node $k$ and 0 if the node is unprotected. We assume that the communication for the node that is both attacked and protected remains intact, i.e. when node $k$ is protected, the attack on this node is not going to succeed. Note that the actual sparsity pattern of the system-wide feedback matrix depends on the actions of both players. Thus, we define the 2-dimensional $N \times N$ array $\bm{S}$ with elements given by vectors of length $n$:
\[\bm{s}_{i,j} = \bm{a}_i \vee \bm{p}_j, \quad i,j = 0,...,N-1 \tag{6} \label{6}\]
where $\vee$ is a bit-wise binary \textit{OR} operator, which for each element $k$ of $\bm{s}_{i,j}$ produces the output 0 only if both inputs are 0, i.e. when the node $k$ is attacked, but not protected. The elements of $\bm{s}_{i,j}$ determine the structural sparsity constraint \cite{ss_Mihailo} on the resulting system-wide feedback matrix $\textbf{K}$, with the sub-blocks $\textbf{K}_{mn}$ in (3) set to zero if $\bm{s}_{i,j}^l = 0$ for $l=m$ and/or $l=n$. 
For example, consider the 3-node network in Fig. 1. The set $\bm{\mathcal{X}}$ contains $N=8$ elements
\[\bm{\mathcal{X}} = \left\{000,001,010,011,100,101,110,111\right\}. \tag{7}\label{7}\]
Suppose the $i=2$ action of the attacker corresponds to disabling communication for nodes 1 and 3 $(\bm{a}_i=(010))$ and the $j=4$ action of the defender is to protect only node 1 $(\bm{p}_j=(100))$. Thus $\bm{s}_{i,j} =(110)$, i.e. for this combination of the attacker's and the defender's actions,  communication is disabled within and to/from node 3, resulting in the sparsity constraint on the feedback matrix $\textbf{K}$, as illustrated in Fig. 2. Note that other combinations of attack and defense actions could also result in the sparsity pattern shown in the figure.
\vspace{-0.2cm} 
\begin{figure}[ht]
\centering
\includegraphics[scale=0.55]{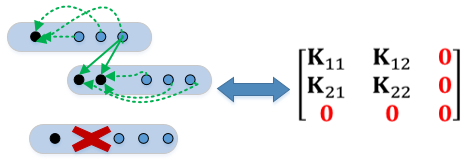}
\caption{\small{Disabled communication for the sparsity pattern  $\bm{s}_{i,j} =(110)$ and the sparsity-constrained feedback matrix resulting from attacking nodes 1 and 3  and defending node 1 only. All communication and local links associated with node 3 are removed.}}
\end{figure}

Next, we define, the energy loss matrix $\bm{\Delta}$ with the elements  
\vspace{-0.3cm}
\[\Delta_{\bm{s}_{i,j}} = J(\textbf{K}^*_{\bm{s}_{i,j}}) - J(\textbf{K}^*_{lqr}), \quad i,j = 0,...,N-1  \tag{8}\label{8}\]
where $\textbf{K}^*_{lqr}$ and $\textbf{K}^*_{\bm{s}_{i,j}}$ optimize  the LQR objective function (4) without and with the structural sparsity constraint imposed by $\bm{s}_{i,j}$, respectively. The second term in (8) indicates the optimal LQR system performance (with unconstrained $\textbf{K}_{lqr}$ matrix, which can have feedback within and among all nodes).  The first term is the smallest energy achieved when the structural sparsity specified by the pattern $\bm{s}_{i,j}$ is imposed on the feedback matrix, computed using the structural sparsity optimization algorithm in \cite{ss_Mihailo}. The loss (8) is the system-wide energy increase caused by the sparsity constraint $\bm{s}_{i,j}$ (i.e the unprotected attacks in the pair $\bm{a}_i, \bm{p}_j$). 

\subsection{The Payoffs and Mixed Strategies}
Given the attacker's and defender's actions $\bm{a}_i$ and $\bm{p}_j$ and the corresponding sparsity pattern $\bm{s}_{i,j}$ (6), the $N \times N$ payoff matrices $\textbf{U}_a$ of the attacker and $\textbf{U}_d$ of the defender have the elements: 
\begin{equation*}
\label{9}\tag{9}
\begin{gathered}
 U_{a_{ij}} = \Delta_{\bm{s}_{i,j}} - \gamma_a(n_{\bm{a}_{i}}), 
 \\
 U_{d_{ij}} = -\Delta_{\bm{s}_{i,j}} - \gamma_d(n_{\bm{p}_{j}}),
 \\
 i,j = 0,...,N-1
\end{gathered}
\end{equation*}
where $\Delta_{\bm{s}_{i,j}}$ is the performance loss (8), $n_{\bm{a}_{i}}$ is the number of zero (attacked nodes) in the pattern $\bm{a}_{i}$, $n_{\bm{p}_{j}}$ is the number of ones (protected nodes) in $\bm{p}_{j}$, and $\gamma_a$, $\gamma_d$ is the player's cost per attacked or defended node, respectively. 

A mixed strategy of the attacker is given by a vector 
\[\bm{r} = (r_0,r_1,...,r_{N-1}), \quad \sum_{i=0}^{N-1} r_i = 1, \tag{10}\label{10}\]
where $r_i \geq 0$ is the probability the attacker chooses to take action $\bm{a}_i$. Similarly, a mixed strategy of the defender is
\[\bm{d} = (d_0,d_1,...,d_{N-1}), \quad \sum_{j=0}^{N-1} d_j = 1, \tag{11}\label{11}\]
where $d_j \geq 0$ is the probability that defender's action is $\bm{p}_j$.
Finally, given the mixed strategies $\bm{r}$ and $\bm{d}$, the expected payoffs of the attacker and defender are given by
\begin{equation}
\label{12}\tag{12}
\begin{aligned}
E_a(\bm{r},\bm{d}) = \bm{r}\textbf{U}_a\bm{d}^T = E(\bm{\Delta}) - E(\bm{C}_a),
\\
E_d(\bm{r},\bm{d}) = \bm{r}\textbf{U}_d\bm{d}^T = -E(\bm{\Delta}) - E(\bm{C}_d),
\end{aligned}
\end{equation}
where \[E(\bm{\Delta}) = \bm{r}\bm{\Delta}\bm{d}^T \label{13}\tag{13}\]
is the expected control performance loss, where $\bm{\Delta}$ is specified by (8), and
\begin{align*}
\label{14-a}\tag{14-a}
E(\bm{C}_a) = \gamma_aE(n_{\bm{a}_{i}}), 
\\
\label{14-b}\tag{14-b}
E(\bm{C}_d) = \gamma_dE(n_{\bm{p}_{j}}), 
\end{align*}
are the expected costs of the attacker and the defender, respectively. From (12), $E_a(\bm{r}^*,\bm{d}^*)\geq 0$ and $E_d(\bm{r}^*,\bm{d}^*) \leq 0$.

The proposed MS game must be played many times to realize the attacker and defender's expected utilities, and it is a one-shot game, i.e., the players do not have information on how each player has acted in the past. In each game realization, the attacker (or defender) aims to maximize its individual payoff $E_a(\bm{r}, \bm{d})$ (or $E_d(\bm{r}, \bm{d})$) by controlling its mixed strategy $\bm{r}$ (or $\bm{d}$), respectively \cite{Wiley-NLP}. To achieve their optimal objectives, the players must balance the costs and control performance terms in (13)-(14), and a higher $\gamma_a$ or $\gamma_d$ value implies that the cost associated with attack or protection, respectively, increases, thus forcing the player to save money and, consequently, compromise its control performance objective. 

\subsection{Computing MSNE}
Since the action spaces of the players are finite (5), 
the proposed game has at least one MSNE \cite{Wiley-NLP} $(\bm{r}^*,\bm{d}^*)$ given by
\begin{align*}
\label{15}\tag{15}
\begin{aligned}
E_a(\bm{r}^*,\bm{d}^*) = \bm{r}^*\textbf{U}_a\bm{d}^{*T} \geq \bm{r}\textbf{U}_a\bm{d}^{*T} = E_a(\bm{r},\bm{d}^*): \forall \bm{r} 
\\
E_d(\bm{r}^*,\bm{d}^*) = \bm{r}^*\textbf{U}_d\bm{d}^{*T} \geq \bm{r}^*\textbf{U}_d\bm{d}^{T} = E_d(\bm{r}^*,\bm{d}): \forall \bm{d} 
\end{aligned}
\end{align*}

The MSNE (15) determines the probabilities the players use to choose their strategies in the game implementation. To compute a MSNE, we employ nonlinear programming (NLP) as follows. The vector $(\bm{r}^*,\bm{d}^*)$ is a MSNE if and only if it satisfies, along with scalars $f^*$, $g^*$, the following nonlinear program, \cite{Wiley-NLP}, Theorem 3.4.1: 
\begin{align*}
\label{16-a}\tag{16-a}
& \hspace{-2.5em} \max_{r_i,d_i,f,g} \
{\sum\limits_{i=0}^{N-1} \sum\limits_{j=0}^{N-1} r_iU_{a_{ij}}d_j + \sum\limits_{i=0}^{N-1} \sum\limits_{j=0}^{N-1} r_iU_{d_{ij}}d_j - f - g}\\
\label{16-b}\tag{16-b}
\mathrm{s.t.} & \hspace{2.5em} \sum\limits_{j=0}^{N-1} U_{a_{ij}}d_j \leq f: \forall i = 0,1,...,N-1 \\
\label{16-c}\tag{16-c}
& \hspace{2.5em} \sum\limits_{i=0}^{N-1} r_iU_{d_{ij}} \leq g: \forall j = 0,1,...,N-1\\
\label{16-d}\tag{16-d}
& \hspace{7em} r_i \geq 0, d_j \geq 0, \\
\label{16-e}\tag{16-e}
& \hspace{5em} \sum\limits_{i=0}^{N-1} r_i = 1, \sum\limits_{j=0}^{N-1} d_j = 1.
\end{align*}
where $f^* = E_a(\bm{r}^*,\bm{d}^*)$ and $g^* = E_d(\bm{r}^*,\bm{d}^*)$. We solve (16) using MATLAB solver \textit{fmincon}.

In summary, a MSNE $(\bm{r}^*,\bm{d}^*)$ computed by the NLP algorithm in (16) determines the players' optimal mixed strategies and the expected utilities of the attacker and the defender (12). These utilities balance the control performance loss (8) and the players' costs. 
Moreover, the optimization (16) guides allocation of security resources by establishing the regions where a desirable level of control performance protection is possible for a given NCS.

\section{Numerical Results}
\subsection{Power system model}
We next show that the proposed MS game can be useful for combating security threats on wide-area control (WAC) loops of power system networks. We analyze the dependency of the game solution on the costs of attack and defense, and illustrate the regions of costs where reliable protection is possible for a given WAC. This investigation
reveals the identity of the important generators for protection in cost-constrained regions. The WAC problem is formulated as follows.

Consider a power system network with $n$ synchronous generators. Each generator is modeled by its \textit{flux-decay} model, which is a common choice for designing wide-area damping controllers using excitation control. The model for the $i^{th}$ generator can be written as
\begin{align*}
\label{17}\tag{17}
\dot{\delta_i} = & \omega_i\\
M_i\dot{\omega_i} = & P_{mi} - d_i \omega_i - \frac{|\bm{V}_i|E_i}{x^{'}_{di}}\sin(\delta_i-\angle \bm{V}_i)  \\ \nonumber
& + \frac{|\bm{V}_i|^{2}}{2}\Big( \frac{1}{x^{'}_{di}} - \frac{1}{x_{qi}} \Big)\sin(2\delta_i-2\angle \bm{V}_i) \tag{18}\label{18}\\
\tau_{doi} \dot{E_i} = &- \frac{x_{di}}{x^{'}_{di}}E_i + \Big(\frac{x_{di}}{x^{'}_{di}} -1  \Big)|\bm{V}_i| \cos(\delta_i-\angle \bm{V}_i) + V_{fdi},\tag{19}\label{19}
\end{align*}
followed by active and reactive power balance equations
\begin{align}
\label{20}\tag{20}
P_i  = &  \frac{E_i|\bm{V}_i|}{x^{'}_{di}} \sin(\delta_i-\angle \bm{V}_i) \\ \nonumber & -\frac{|\bm{V}_i|^{2}}{2}\Big( \frac{1}{x^{'}_{di}} - \frac{1}{x_{qi}} \Big)\sin(2\delta_i-2\angle \bm{V}_i)\\
\label{21}\tag{21}
Q_i = &\frac{E_i|\bm{V}_i|}{x^{'}_{di}} \cos(\delta_i-\angle \bm{V}_i) \\
&- |\bm{V}_i|^{2} \Bigg(\frac{\sin^{2}(\delta_i-\angle \bm{V}_i)}{x_{qi}}  -  \frac{\cos^{2}(\delta_i-\angle \bm{V}_i)}{x^{'}_{di}} \Bigg). \nonumber
\end{align}

\noindent Equations (17)-(18) represent the electro-mechanical swing dynamics, and (19) represents the electro-magnetic dynamics of the $i^{th}$ generator. $\delta_i$ is the rotor phase angle, $\omega_i$ is the rotor speed, $E_i$ is the internal voltage of the $i^{th}$ generator, while  $\bm{V}_i$ is the voltage phasor at the generator bus, $P_i$ and $Q_i$ are the active and reactive power outputs of the generator, and $V_{fdi}$ is the exciter voltage. Definitions of the various model parameters are standard in the literature \cite{kundur}. The generator model is coupled with the model of an exciter consisting of an automatic voltage regulator (AVR) and a power system stabilizer (PSS) whose combined dynamics can be written as
\begin{align*}
\label{22}\tag{22}
 \tau_{e_i} \dot{V}_{fdi} = & -V_{fdi} + V^{\bigstar}_{fdi} + K_{ai}( |\bm{V}_i| - |\bm{V}_i|^{\bigstar} - \nu_i + \gamma_i )\\
\label{23}\tag{23}
 \dot{\zeta_i} = & \bm{A}_{pss}\zeta_i + \bm{B}_{pss}\omega_i,\mbox{ } \nu_i = C_{pss}\zeta_i + D_{pss}\omega_i
\end{align*}
\noindent where superscript $^{\bigstar}$ means set-point. The signal $\gamma_i$ serves as a control input representing an additional voltage reference signal to the AVR that can be designed to add damping to the  slow  or  inter-area  oscillation  modes using state feedback from all generators spread across the grid. These controllers are referred to as wide-area controllers (WAC).

Our control design does not necessarily need the generators to follow this simple model. Detailed models of generators are allowed, provided all the generator states can be measured or estimated (for details of decentralized phasor-based state estimation please see \cite{pal}). 
In general, we assume the $i^{th}$ generator to consist of $m_i$ states $\bm{\xi}_i = [\delta_i, \; \omega_i, \; \bm{x}_{i,rem}] \in \mathbb R^{m_i}$ where $\bm{x}_{i,rem}$ is the vector of all non-electromechanical states, and one scalar control input $\gamma_{i}$ as in (22), which is the field excitatin voltage, and $\sum_{i=1}^{n}m_i = m$ (total number of states).
Let the pre-disturbance equilibrium of the $i^{th}$ generator be $\xi^{*}_i=[\delta^{*}_i, \; \omega^{*}_i, \; \bm{x}_{i,rem}^{*}]$.  
The differential-algebraic model of the generators and the power flow is converted to a state-space model using Kron reduction \cite{kundur}, and linearized about $\xi_i^{*}, i = 1,2,...,n$. 
The small-signal model of the system with the $i^{th}$ state defined as $\bm{x}_i = \xi_i - \xi_i^{*}$, can then be written as (1). Note that the state vector $\bm{x}_i$ includes the AVR and PSS states from (22)-(23) linearized around their respective equilibria. The small-signal control input is given by
$u_i = \Delta \gamma_i$, and the disturbance $w$ in (1) is modeled as an impulse input entering through the acceleration equation of the generator. 

We model the attacks where the wide-area communication to/from and within a generator (node) is disabled when a hardware attack occurs, keeping the PSS intact for all generators.
The matrix $\textbf{R}$ is chosen as the identity matrix, while $\textbf{Q}$ is chosen such that all the generators arrive at a consensus in their small-signal changes in phase angle and frequencies, as dictated by the physical topology of the network \cite{Feier}. The objective function is chosen as 
\[E_{states} = \Delta\delta^T\bar{\mathcal{L}}\Delta\delta + \Delta\omega^T\bar{\mathcal{L}}\Delta\omega + \Delta\bm{x}_{rem}^T\bar{\mathcal{L}}\Delta\bm{x}_{rem} \label{24}\tag{24}\]
where $\bar{\mathcal{L}}$ is constructed as
\[\bar{\mathcal{L}} = m\textbf{I} - \textbf{1}_m.\textbf{1}_m^T \tag{25}\label{25}\]
where \textit{m} is the total number of states. Physically, this means that the first two terms in the objective function (24) are in the consensus form
\begin{equation}
\tag{26}\label{26}
\begin{aligned}
\Delta\delta^T\bar{\mathcal{L}}\Delta\delta = \sum_{i=1}^{m} \sum_{j>i}^{m} (\Delta\delta_i -\Delta\delta_j)^2.
\\
\Delta\omega^T\bar{\mathcal{L}}\Delta\omega = \sum_{i=1}^{m} \sum_{j>i}^{m} (\Delta\omega_i - \Delta\omega_j)^2. 
\end{aligned}
\end{equation}
Then (26) can be expressed as
\begin{gather*}
\tag{27}\label{27}
E_{states} =
\begin{bmatrix}
\Delta\delta \\
\Delta\omega\\
\Delta\bm{x}_{rem}
\end{bmatrix}^T
\underbrace{
\begin{bmatrix}
\bar{\mathcal{L}} & & \\
& \bar{\mathcal{L}} & \\
& & \textbf{I}
\end{bmatrix}}_{\textbf{Q}}
\begin{bmatrix}
\Delta\delta \\
\Delta\omega\\
\Delta\bm{x}_{rem}
\end{bmatrix}\\
= \bm{x}^T(\textbf{Q}) \bm{x}. 
\end{gather*}

\subsection{Game results for the New England Power System }
Next, we employ the WAC characterization (17-27) to investigate the proposed game for the IEEE 39-bus New England power system model. The model consists of 10 synchronous generators spread among 39 buses. Generator 1 is modeled by 7 states, generators 2 to 9 are modeled by 8 states each, and generator 10 by 4 states.
\begin{table}[h]
\label{Table: Strategies dependent on cost parameters}
\begin{center}
\resizebox{\columnwidth}{!}{%
\begin{tabular}[c]{ | P{2cm} | P{2cm} | P{2cm} |} 
\hline

\rule{0pt}{10pt}
\bf{Disabled generator subset} & \bf{Fract. loss $\%$  (local links disabled)} & \bf{Fract. loss $\%$ (local links intact}) \\[2pt]
\hline 
\hline 

2 & 4.91 &  0.03  \\[2pt]
\hline 

4 & 5.32 & 0.06 \\[2pt]
\hline 

6 & 5.01 & 0.05 \\[2pt]
\hline 

8 & 2.57 & 0.02\\[2pt]
\hline 

1,7,8 & 10.23 & 0.10 \\[2pt]
\hline 

9 & 6.37 & 0.09\\[2pt]
\hline

2,4,6,9 & 34.31 & 0.23 \\[2pt]
\hline

1,2,3,4,5,6,7,8,9 & $2.9\times 10^5$ (open-loop) & 0.5  \\[2pt]
\hline

\end{tabular}
}
\end{center}
\caption{\small{Disabled generator subsets for sparsity patterns $\bm{s}_{i,j}$ (6) with dominant fractional losses  $\frac{\Delta_{\bm{s}_{i,j}}}{J(\textbf{K}^{*}_{lqr})} (\%)$}}
\end{table}

Table 1 illustrates the fractional control performance losses given by $\Delta_{\bm{s}_{i,j}}(\%)$ (see (8)) for the sparsity  
patterns $\bm{S}$ (6) with $J(\textbf{K}^*_{lqr})$. Note that when generators $9,4,6,2$ are disabled jointly, high loss occurs. However attacking or protecting this set also incurs high cost for the players. Note that for patterns where only single unprotected generator is attacked, the highest loss is observed for the disabled generator $9$, 
followed by generators $4,6,2,3,1,5,7,8$, imposing the importance order of the nodes. 

From Table 1, we observe that when only the inter-node communication links are disabled while the self-links are intact (see Fig. 1), the losses are greatly reduced, confirming that the self-links are critical to the system's control performance. The \textit{inter-node} communication links can be compromised using cyber-attacks in the communication medium. They are less expensive than node attacks investigated in this paper, which require hardware manipulation, but also affect the self-links and thus lead to much stronger damage as shown in Table 1. Note that when WAC communication between and within all generators is disabled, extremely large loss results since the system is in 
open-loop operation (see last entry of Table 1). On the other hand, when self-links are intact, but the generators cannot communicate with each other, the loss is relatively low. 
Note that the papers on structural sparsity \cite{ss_Mihailo}, sparsity promotion \cite{Mihailo} and sparsity-constrained \cite{Feier} algorithms did not include examples where the self-links were disabled. However, the algorithms in these papers can be extended to the latter case. For example, when in the sparsity-promotion Alternating Direction Method of Multipliers (ADMM) method in \cite{Mihailo}, the parameter $\gamma$ is increased beyond the range studied in that paper, the self-links are removed in the order that corresponds to the order of the losses illustrated in Table 1, thus confirming the decreasing order of importance of generators $9,4,6,2$, as well as the significance of self-links shown in Table 1. In the rest of the paper, we assume that when unprotected nodes are attacked, the self-links are disabled, thus reflecting the losses associated with hardware-level attacks.

In Fig. 3(a)-(b), we illustrate the expected payoffs of the players (12), the expected fractional loss, and the expected fractional costs $\frac{E(\bm{C}_a)}{J(\textbf{K}^{*}_{lqr})} (\%)$, and $\frac{E(\bm{C}_d)}{J(\textbf{K}^{*}_{lqr})} (\%)$ (14), at MSNE (15) vs. $\gamma_d$ when $\gamma_a = 0.01$.
Moreover, the attacked/protected node subsets with dominant probabilities at MSNE are shown in Fig. 3(c) for some $\gamma_d$ values.
Note that as $\gamma_d$ increases, the number of protected nodes decreases, thus reducing and increasing the payoffs of the defender and the attacker, respectively. 
The expected control performance loss (13) also grows as defense investment decreases for higher values of $\gamma_d$ since attacks become more successful. Note that the expected cost of the defender (14-b) also grows with $\gamma_d$, demonstrating the control performance/cost trade-off. On the other hand, the expected cost of the attacker (14-a) saturates at a low value (14.19 $\%$). Since $\gamma_a$ is small, the attacker can afford to attack all nodes with high probability as shown in Fig. 3(c). However, as $\gamma_a$ grows, the attacker's ability to cause significant damage quickly reduces as will be illustrated below.

In Fig. 3(c), when $\gamma_d = 0$, a pure-strategy NE occurs since the defender protects all nodes, and the attacker saves money
\begin{figure}[H]
  \begin{center}
    \begin{minipage}{0.46\textwidth}
      \includegraphics[width=\textwidth]{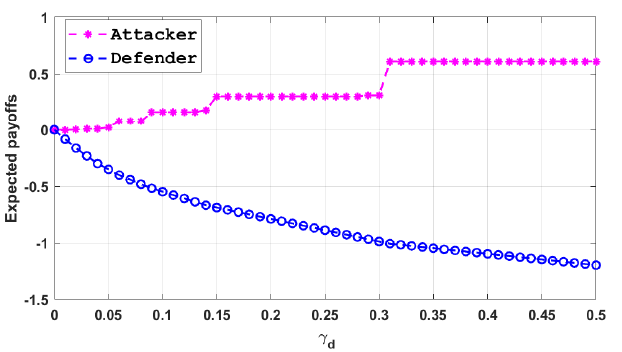}
      \caption*{(a)}
       \end{minipage}%
       
       \begin{minipage}{0.46\textwidth}
       \includegraphics[width=\textwidth]{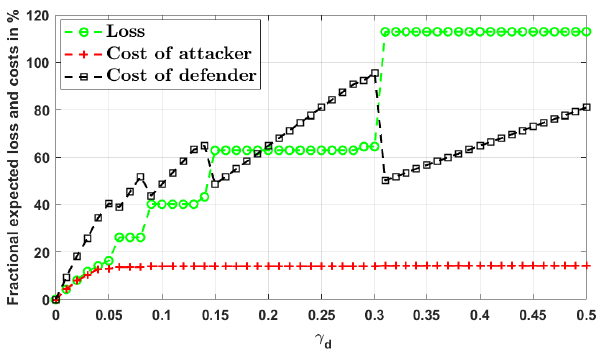}
       \caption*{(b)} 
      \end{minipage}%
      
      \begin{minipage}{0.47\textwidth}
      \includegraphics[width=\textwidth]{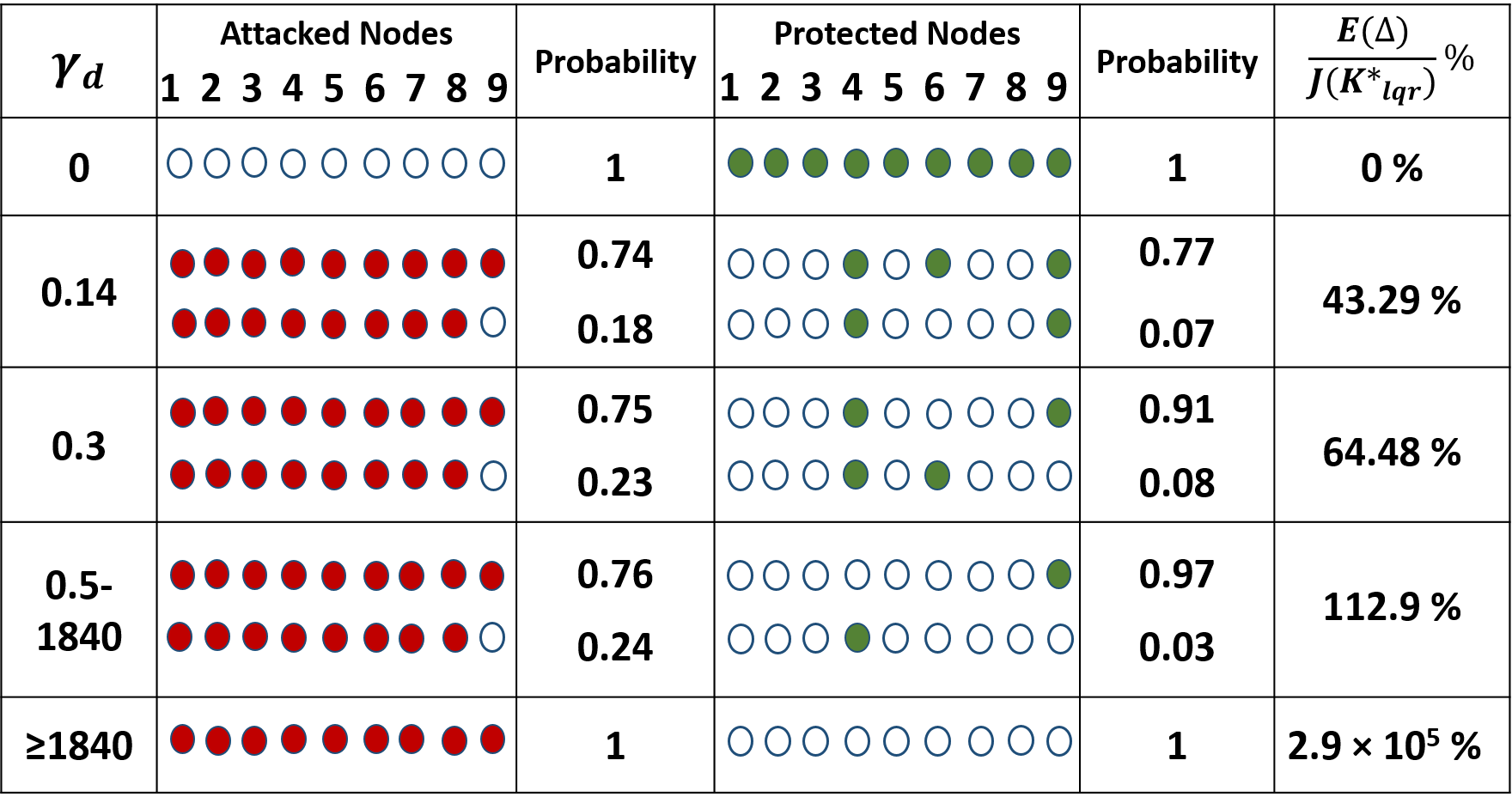}
      \caption*{(c)}
      \end{minipage}
\caption{\small{Performance at MSNE vs. $\gamma_d$; $\gamma_a = 0.01$; (a) Expected payoffs of the players (12); (b) Fractional expected energy loss $\frac{E(\bm{\Delta})}{J(\textbf{K}^{*}_{lqr})}$ and costs $\frac{E(\bm{C}_a)}{J(\textbf{K}^{*}_{lqr})}$ and $\frac{E(\bm{C}_d)}{J(\textbf{K}^{*}_{lqr})}$ $(\%)$; (c) Attacked/protected subsets of nodes with two highest probabilities (all other probabilities are below 0.03) and expected fractional loss $\frac{E(\bm{\Delta})}{J(\textbf{K}^{*}_{lqr})} (\%)$}}
   \end{center}
\end{figure}
\vspace{-0.21cm}
\noindent by not acting at all. 
Note that other (mixed) NEs are also possible, but the payoff is zero for both players for all these NEs, implying that the control performance is preserved. As $\gamma_d$ increases, the defender invests in the more important subsets of nodes while the attacker invests mostly into attacking all nodes due to low $\gamma_a$ value.
The transitions in the curves in Fig. 3(a)-(b) correspond to the reduction in the numbers of protected nodes as  
$\gamma_d$ grows, which is illustrated by node subsets in Fig. 3(c). As $\gamma_d$            approaches 0.3, the 
defender mostly protects the important nodes 9 and 4, and for $0.5 \leq \gamma_d < 1840$, the most important generator $9$ is protected with very high probability due to high cost of defense per node. The corresponding control performance loss grows as $\gamma_d$ increases. 
Finally, for very high $\gamma_d \geq 1840$, 
we obtain a pure NE where the defender chooses not to protect any nodes and, since $\gamma_a$ is very low, the attacker successfully disables all communication in the system, resulting in the open-loop operation, which has the total loss of $1840$ or the fractional loss $2.9 \times 10^5 \%$.
\begin{figure}[t]
  \begin{center}
    \begin{minipage}{0.455\textwidth}
      \includegraphics[width=\textwidth]{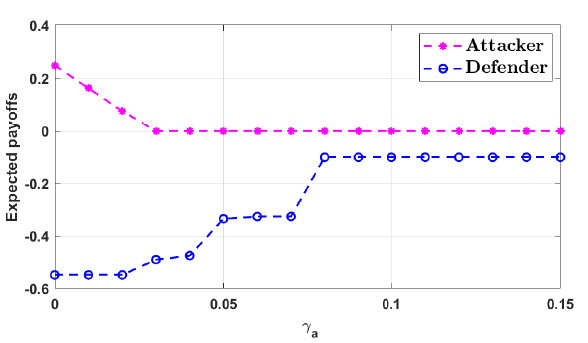}
      \caption*{(a)}
       \end{minipage}%
       
              \begin{minipage}{0.47\textwidth}
       \includegraphics[width=\textwidth]{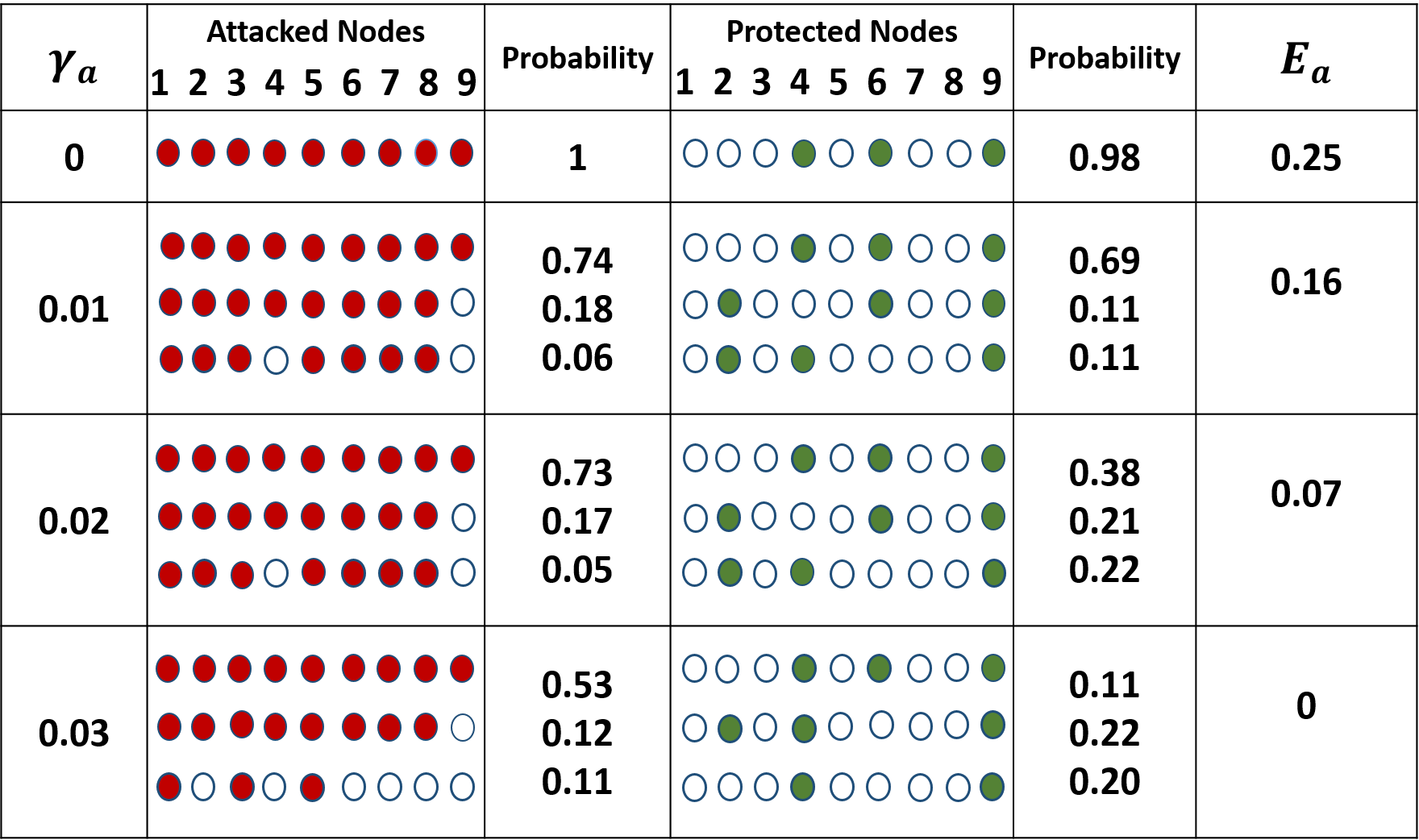}
       \caption*{(b)} 
      \end{minipage}
\caption{\small{Performance at MSNE vs. $\gamma_a$; $\gamma_d = 0.1$; (a) Expected payoffs of the players (12); (b) Attacked/protected subsets of nodes with highest probabilities (all other probabilities are below 0.03) and the attacker's expected utility $E_a(\bm{r}^*,\bm{d}^*)$}}
   \end{center}
\end{figure}

In Fig. 4, we illustrate the expected payoffs of the players  at MSNE vs. $\gamma_a$ for $\gamma_d = 0.1$. Fig. 4(b) shows the attacked/protected node subsets with the dominant probabilities at MSNE for some values of $\gamma_a$. We observe that the defender protects the important nodes while the attacker starts allocating resources to the unprotected nodes as $\gamma_a$ increases. Note that for $\gamma_a \geq 0.03$, the attacker's payoff is zero. Thus, it is not advantageous for the attacker to act in this region, resulting in successful system protection.

\begin{figure}[h]
  \begin{center}
    \begin{minipage}{0.45\textwidth}
      \includegraphics[width=\textwidth]{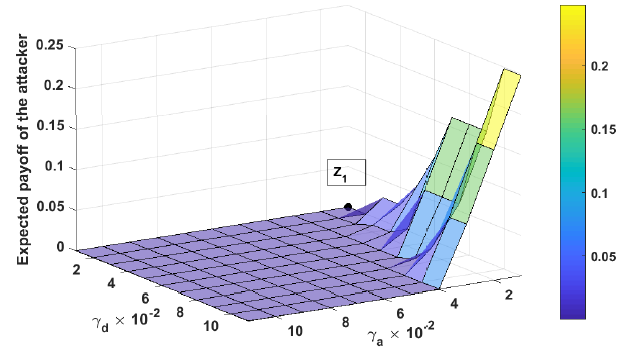}
      \caption*{(a)}
       \end{minipage}%
 
       \begin{minipage}{0.445\textwidth}
       \includegraphics[width=\textwidth]{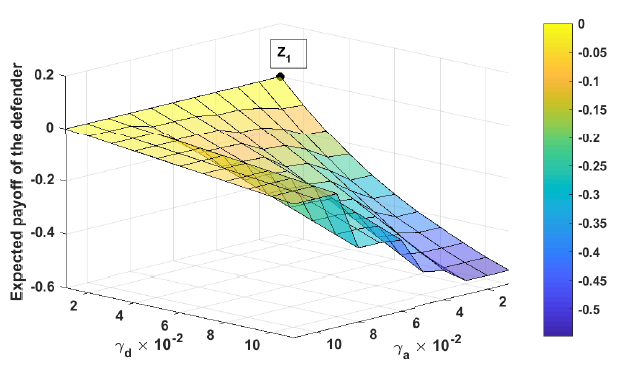}
       \caption*{(b)} 
      \end{minipage}
\caption{\small{Game performance at MSNE vs. $\gamma_a$ and $\gamma_d$; The expected  players' payoffs (12); (a) Attacker; (b) Defender}}
   \end{center}
\end{figure}

Fig. 5 illustrates the dependency of the expected payoffs of the players on the costs $\gamma_a$ and $\gamma_d$, with values in the interval $(0,0.1)$. First, we consider the special case when $\gamma_a = \gamma_d = 0$ in (9), i.e., the costs are negligible, or both players have unlimited resources. On solving the zero-sum game, we obtain a pure NE with all nodes attacked and protected with probability 1, respectively, which results in zero expected payoffs of both players and thus effective system protection, as shown by point $Z_1$. As $\gamma_d$ increases, the defender's payoff diminishes, consistent with Fig. 3 results. We also observe that unless $\gamma_a$ is very small (much smaller than the expected loss) or $\gamma_d >> \gamma_a$, the attacker's payoff at MSNE (15) is $E_a(\bm{r}^*,\bm{d}^*) = 0$. Since the attacker can obtain this payoff by not acting, it loses its incentive to attack the system, resulting in successful protection for most of the region shown in Fig. 5. (Note that the game can have multiple MSNEs (15) \cite{Wiley-NLP}, but a MSNE where the attacker does not act with probability 1 is guaranteed to exist for all $(\gamma_a,\gamma_d)$ points where $E_a(\bm{r}^*,\bm{d}^*) = 0$ in Fig. 4 and 5).

Finally, the routine for finding a MSNE for a given set of $(\gamma_a,\gamma_d)$ values runs in under $900$ seconds\footnote{The experiments are run using MATLAB on Windows 10 with 64-bit operating system, 3.4 GHz Intel core i7 processor, and 8GB memory}. The computational load of the structural sparsity optimization \cite{ss_Mihailo} for each sparsity pattern $\bm{s}_{i,j}$ (6) dominates the overall runtime of the game implementation, but this computation can be shared for all $(\gamma_a,\gamma_d)$ points. We can limit the total time of the latter optimization by limiting the action space (5) to sets of important nodes (which can be found using \cite{Mihailo} or \cite{Feier}). The NLP algorithm (16) is also computationally intensive due to the large size of the system matrices. Since the proposed game solves a resource-planning problem and is implemented offline, the computation complexity does not significantly impact its implementation.

\section{Conclusions}
We developed a mixed-strategy investment game between an attacker and a defender of a network control system. The proposed game allows to allocate the players' resources strategically to save costs and optimize their control performance objectives. 
We found that 
the defender can successfully protect the network from hardware-level attacks that disable communication and thus compromise the control performance unless the cost of attack is very low or is significantly exceeded by the cost of protection per system node. Moreover, as the cost of defense per node increases, a MSNE of the game reveals the important nodes that must be protected to avoid large control performance loss. Future research will address practical game implementation under limited system knowledge by the players.

\balance



%
%
%
%
%
%
%


\end{document}